\documentclass[aps,twocolumn,showpacs,superscriptaddress]{revtex4}
\usepackage[dvips]{graphicx}
\usepackage{amsmath}
\usepackage{amsfonts}
\usepackage{amssymb}
\usepackage[linktocpage,breaklinks=true,colorlinks=true,linkcolor=black,filecolor=black,citecolor=black,pagecolor=black,urlcolor=black]{hyperref}

\newcommand{\ie}{i.e.\ }

\newcommand{\Tr}{\operatorname{Tr}}
\newcommand{\ii}{\mathrm{i}}
\newcommand{\var}{\mathrm{Var}}
\newcommand{\cov}{\mathrm{Cov}}
\newcommand{\ket}[1]{\left|{#1}\right\rangle}
\newcommand{\bra}[1]{\left\langle{#1}\right|}
\newcommand{\ketbra}[2]{\left|{#1}\middle\rangle\middle\langle{#2}\right|}
\newcommand{\proj}[1]{\ketbra{#1}{#1}}
\newcommand{\ew}[1]{\left\langle{#1}\right\rangle}

\newcommand{\DC}{\Delta_\mathrm{c}}
\newcommand{\omrec}{\omega_\mathrm{R}}
\newcommand{\Erec}{E_\mathrm{R}}
\newcommand{\alphac}{\alpha_\mathrm{c}}
\newcommand{\Cp}{\mathcal{C}_p}
\newcommand{\EN}{\mathcal E_\mathcal{N}}

\begin{document}

\title{Quantum-correlated motion and heralded entanglement of distant optomechanically coupled objects}

\author{Wolfgang Niedenzu}
\affiliation{Institut f{\"u}r Theoretische Physik, Universit{\"a}t Innsbruck, Technikerstra{\ss}e~25, 6020~Innsbruck, Austria}

\author{Raimar M. Sandner}
\affiliation{Institut f{\"u}r Theoretische Physik, Universit{\"a}t Innsbruck, Technikerstra{\ss}e~25, 6020~Innsbruck, Austria}

\author{Claudiu Genes}
\affiliation{Institut f{\"u}r Theoretische Physik, Universit{\"a}t Innsbruck, Technikerstra{\ss}e~25, 6020~Innsbruck, Austria}
\affiliation{Vienna Center for Quantum Science and Technology (VCQ), Faculty of Physics, University of Vienna, Boltzmanngasse~5, 1090~Vienna, Austria}
\affiliation{ISIS (UMR 7006) and IPCMS (UMR 7504), Universit\'e de Strasbourg and CNRS, Strasbourg, France}

\author{Helmut Ritsch}
\email{Helmut.Ritsch@uibk.ac.at}
\affiliation{Institut f{\"u}r Theoretische Physik, Universit{\"a}t Innsbruck, Technikerstra{\ss}e~25, 6020~Innsbruck, Austria}

\begin{abstract}
  The motion of two distant trapped particles or mechanical oscillators can be strongly coupled by light modes in a high finesse optical resonator. In a two mode ring cavity geometry, trapping, cooling and coupling is implemented by the same modes. While the cosine mode provides for trapping, the sine mode facilitates ground state cooling and mediates non-local interactions. For classical point particles the centre-of-mass mode is strongly damped and the individual momenta get anti-correlated. Surprisingly, quantum fluctuations induce the opposite effect of positively-correlated particle motion, which close to zero temperature generates entanglement. The non-classical correlations and entanglement are dissipation-induced and particularly strong after detection of a scattered photon in the sine mode. This allows for heralded entanglement by post-selection. Entanglement is concurrent with squeezing of the particle distance and relative momenta while the centre-of-mass observables acquires larger uncertainties.
\end{abstract}

\date{\today}


\maketitle

\section{Introduction}

The past decades have seen tremendous success in the implementation of control schemes for the motional state of matter via light fields either in free space or in optical cavities. A diversity of examples exist where the quantum regime of motion has been reached. The masses span many orders of magnitude, from the microscopic atomic size systems such as atoms in optical cavities~\cite{domokos2003mechanical,mekhov2012quantum,gupta2007cavity,leroux2010implementation} and laser-cooled ions in ion traps~\cite{eschner2003laser} to the macroscopic level with cavity-embedded membranes~\cite{thompson2008strong}, mirrors~\cite{groeblacher2009demonstration} or levitated dielectric nano-particles~\cite{romero2011optically}.
\par
A common interaction Hamiltonian that well approximates many quantum light--matter interfaces is quadrature--quadrature coupling~\cite{wallquist2010single}; more specifically, the displacement of the mechanics is coupled directly to a quadrature of the high-$Q$ optical field mode that can be then used as an observable for indirect position detection. Adding a second mechanical system coupled to the field then allows one to engineer an effective two-particle mechanical coupling by eliminating the mediating light mode. Recently, an expansion to quadratic coupling has been proposed~\cite{nunnenkamp2010cooling} and the investigation of dissipation-induced~\cite{krauter2011entanglement,nha2004entanglement,plenio1999cavity}, noise-induced~\cite{abdi2011effect} and remote entanglement~\cite{boerkje2011proposal,joshi2012entanglement} has been of great interest, including a scheme for sensitive force measurements~\cite{mancini2003high} and entanglement of macroscopic oscillators~\cite{mancini2002entangling,hartmann2008steady}.
\par
Here we show that all this can be implemented in a system consisting of two particles strongly trapped in the cosine mode of a ring cavity, where the two-particle interaction is carried by sideband photons in the sine mode. For deep trapping it yields the typical linearized optomechanical Hamiltonian~\cite{schulze2010optomechanical}. First we present the general model of two particles moving within a ring resonator. We then analyse single quantum trajectories depicting strong correlations and entanglement. A subsequent investigation of momentum correlations reveals classically forbidden positive values in steady state, even in the absence of entanglement. The steady state shows a strong delocalization of the centre-of-mass independent of the particle separation. We also show how to generate entanglement either in a pulsed regime or heralded by the detection of photons. Analytical calculations are carried out in the regime of strong particle confinement and matched to the more generally valid numerical simulations with good agreement. Finally, the occurrence of correlations in the system is explained in a simple adiabatic model.
\par
We structure this presentation in two major parts according to the Gaussian/non-Gaussian nature of the system dynamics. Entanglement on trajectories and that on average as well as momentum correlations during the non-Gaussian dynamics of the two atoms in the cavity are numerically investigated in the beginning, while the Gaussian approximation makes the topic of the second major part of the paper. This part corresponds either to atoms in very deep traps (to compare to the results obtained in the first part) or more generally to a wide range of optomechanical systems in the linear regime and with diverse physical realizations (moving membranes, levitated nano-spheres etc.).

\section{Motion of two particles in a ring resonator}

\begin{figure}
  \centering
  \includegraphics[width=0.6\columnwidth]{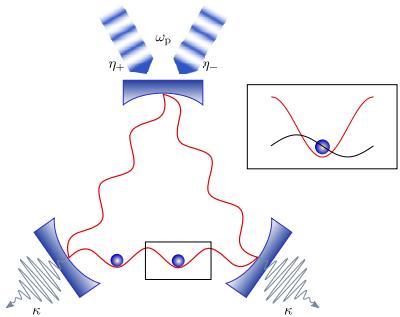}
  \caption{Sketch of the system. The pumped standing-wave mode (red) traps the particles, the second orthogonal one (black) mediates an effective interaction between them.}\label{fig_resonator}
\end{figure}

We study two small polarizable particles confined within a symmetrically-pumped ring resonator, see figure~\ref{fig_resonator}. Symmetric pumping results in a standing-wave optical potential with spatial dependence $\cos^2(kx)$~\cite{schulze2010optomechanical,niedenzu2010microscopic}. The cosine mode is strongly pumped and approximated by a highly excited coherent state $\ket\alphac$ with $|\alphac|\gg1$ (and without loss of generality $\alphac\in\mathbb{R}$). The particles scatter photons into the unpumped orthogonal sine mode. This setup can be generally described by the Hamiltonian~\cite{schulze2010optomechanical,niedenzu2010microscopic}
\begin{multline}\label{eq_H}
  H=\sum_{i=1}^2\left[\frac{p_i^2}{2m}+\hbar U_0\alphac^2\cos^2(kx_i)+\hbar U_0 a^\dagger a\sin^2(kx_i)\right]+\\
  +\frac{\hbar U_0\alphac}{2}\left(a+a^\dagger\right)\sum_{i=1}^2\sin(2kx_i)-\hbar\DC a^\dagger a.
\end{multline}
Here, $a$ denotes the annihilation operator of the quantum-mechanical sine mode, $U_0<0$ the optical potential depth per photon, $x_i$ and $p_i$ the particles' centre-of-mass position and momentum operators, respectively, and $m$ the particle mass. The pump is detuned by $\DC:=\omega_\mathrm{p}-\omega_\mathrm{c}$ from the bare cavity resonance frequency $\omega_\mathrm{c}$. To avoid instabilities we restrict ourselves to red detuned lasers ($\DC<0$) for which a cooling regime exists~\cite{gangl2000cold}. The sine mode is only weakly populated by scattering such that $\ew{a^\dagger a}$ is negligible compared to $\alphac^2$~\cite{schulze2010optomechanical}. Damping of the cavity mode is taken into account by the Liouvillian $\mathcal L\rho=\kappa\left(2a\rho a^\dagger - a^\dagger a\rho - \rho a^\dagger a \right)$~\cite{gardinerbook} in the master equation
\begin{equation}\label{eq_master}
  \dot\rho=\frac{1}{\ii\hbar}\left[H,\rho\right]+\mathcal{L}\rho.
\end{equation}

To get some insight into the dynamics, we solve equation~\eqref{eq_master} numerically for some typical parameters. The direct solution of the master equation~\eqref{eq_master} is computationally very demanding owing to the large Hilbert space of the joint particles-field system. Therefore, we resort to Monte Carlo wavefunction simulations~\cite{moelmer1993monte}, in which the system is coherently evolved between the so-called ``quantum jumps''. These jumps correspond to a photon detected at the resonator output~\cite{moelmer1993monte}. Besides the more favourable usage of computer resources, single trajectories also provide an additional insight into the microscopic processes in the system. The simulations were efficiently implemented with the freely available\footnote{\url{http://cppqed.sourceforge.net}} C++QED framework~\cite{vukics2007object,vukics2012cppqedv2} and performed in a joint momentum- and Fock basis. 
\par
\begin{figure}
  \centering
  \includegraphics[width=0.8\columnwidth]{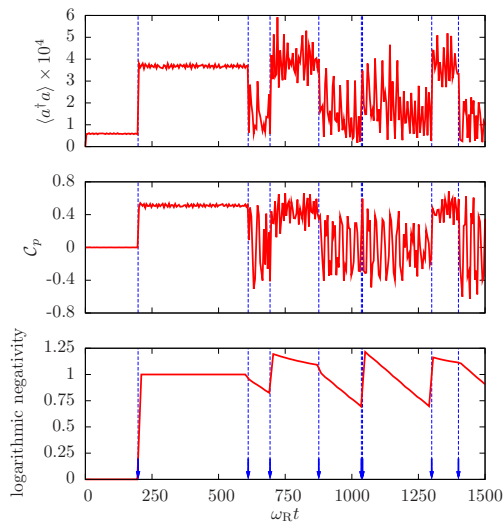}
  \caption{Photon number, momentum correlation and entanglement for a single trajectory. The dashed vertical lines indicate quantum jumps. Parameters: $\alphac=150$, $U_0=-\frac{1}{\alphac}\omrec$, $\DC=-\kappa$ and $\kappa=10\omrec$.}\label{fig_singletrajectory}
\end{figure}

A typical trajectory is shown in figure~\ref{fig_singletrajectory}, where the blue arrows indicate the times at which jumps occur. Initially, the particles were prepared in the ground state of two separated potential wells and owing to the deep potential tunnelling is strongly suppressed. The momentum correlation coefficient $\Cp$ for the two particles is defined as
\begin{equation}\label{eq_Cp}
  \Cp:=\frac{\cov(p_1,p_2)}{\Delta p_1\Delta p_2}=\frac{\ew{p_1p_2}-\ew{p_1}\ew{p_2}}{\Delta p_1\Delta p_2},
\end{equation}
where $\Cp=1$ means perfect correlation and $\Cp=-1$ perfect anti-correlation of the motion. Quantum jumps by photodetection trigger strong correlations and entanglement between the particles due to the cavity-mediated interaction. The logarithmic negativity~\cite{vidal2002computable} already after the first jump approaches the value for a maximally entangled Bell state and the correlation reaches a value of $\Cp\approx 0.5$. The emerging state corresponds to a superposition of two particles moving to the right and two particles moving to the left such that the centre-of-mass momentum remains zero. As we will see later this behaviour is caused by the excitation of a single particle into the first excited state within the trap. Beginning with the second jump the system is subject to fast oscillations which are of the order of the trap frequency. Note that only the field is dissipative in our model and induces quantum jumps. However, the particles respond to the sudden changes of the field in a correlated way.

\par
Interestingly, while entanglement is quite pronounced on single trajectories, it remains small on average. The averaged steady-state logarithmic negativity is $\EN\sim10^{-3}$ for our parameters. The logarithmic negativity is not an expectation value and hence its value for an ensemble of trajectories cannot be obtained from its values on single realizations---it has to be directly computed from the density operator. It is therefore possible that the logarithmic negativity of a mixture of entangled states is smaller than the weighted average of the individual values. A prominent example for this is the equal statistical mixture of two Bell states, which is not entangled.

\begin{figure}
  \centering
  \includegraphics[width=0.8\columnwidth]{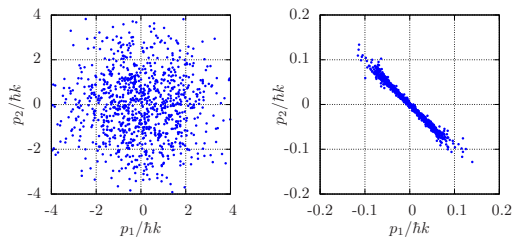}
  \caption{Classical simulations~\cite{gangl2000cold} for 1000 initial conditions. Left: initial condition, right: distribution at $\omrec t=3000$. The particles become anti-correlated and cooled. Same parameters as in figure~\ref{fig_singletrajectory}.}\label{fig_classical}
\end{figure}

\begin{figure}
  \centering
  \includegraphics[width=0.8\columnwidth]{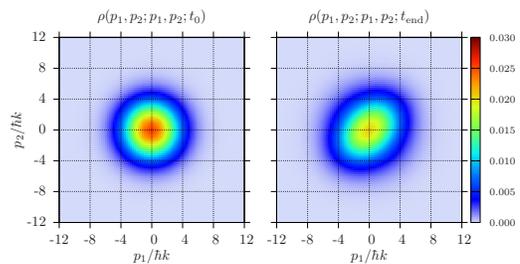}
  \caption{Diagonal elements of the two-particle reduced density matrix in momentum space, initially (left) and in steady state (right). Ensemble average of 5000 trajectories. Same parameters as in figure~\ref{fig_singletrajectory}.}\label{fig_rho}
\end{figure}

\begin{figure}
  \centering
  \includegraphics[width=0.8\columnwidth]{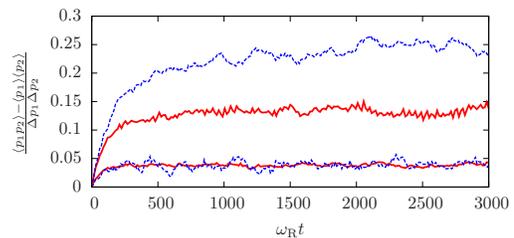}
  \caption{Red solid lines: momentum correlation coefficient obtained from the master equation containing~\eqref{eq_H} for $\DC=-\kappa$ (upper curve) and $\DC=-\omega$ (lower curve). Blue dashed lines: result of the oscillator approximation~\eqref{eq_H_ho} (600 trajectories). The other parameters are the same as in figure~\ref{fig_singletrajectory}.}\label{fig_momentumcorrelation}
\end{figure}

We now examine the momentum correlations in more detail. Classical simulations reveal much stronger damping of the centre-of-mass motion than that of the relative motion. Hence the particles become anti-correlated~\cite{gangl2000cold,gangl2000cooling,domokos2003mechanical}, see figure~\ref{fig_classical}. Surprisingly, the quantum simulations of the ring resonator system yield the opposite result. On average, initially uncorrelated particles become positively correlated due to the cavity input noise. This effect is visible in the steady-state density matrix presented in figure~\ref{fig_rho}. There, the momentum distribution is elongated into the first quadrant, which is a signature of positive correlations. Note that quantum mechanics allows pure states with positive correlations, but still zero average centre-of-mass motion with large uncertainty. The time evolution of the momentum correlation coefficient is depicted in figure~\ref{fig_momentumcorrelation}. We observe smaller correlations around the cooling sideband $\DC=-\omega:=-\sqrt{4\hbar|U_0|\alphac^2\Erec}/\hbar$ as compared to the other chosen detuning $\DC=-\kappa$. Here $\Erec\equiv\hbar\omrec:=\hbar^2k^2/2m$ is the recoil energy. For comparison, we also present the results of simulations containing the quadratic optomechanical Hamiltonian~\eqref{eq_H_ho} introduced later on in this paper. The latter yields accurate results provided that the position spread remains small compared to a wavelength, $k\Delta x\ll 1$. This is fulfilled for operation near the cooling sideband, but not for $\DC=-\kappa$. Positive correlations of the particle motion can also be observed in much shallower potentials where the particles are barely trapped and both light modes are treated quantum-mechanically~\cite{sandner2012}.

\begin{figure}
  \centering
  \includegraphics[width=0.8\columnwidth]{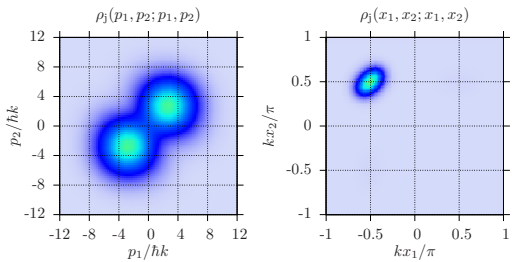}
  \caption{Diagonal elements of the reduced two-particle conditional density matrix~\eqref{eq_rhoj} after a photon detection in steady state. Particle correlations as well as entanglement become much more pronounced compared to the density matrix shown in figure~\ref{fig_rho}. Same parameters and colour code as there.}\label{fig_rhoj}
\end{figure}

In the dynamics, correlations and entanglement are generated by quantum jumps induced by photon count events. Hence they are particularly strong immediately after a jump. In real experiments, it is generally not possible to exactly keep track of each emitted photon due to the detector efficiency, \ie one cannot exactly follow a certain quantum trajectory. Hence, the system always evolves into a mixed state and if a photon is observed (measured) outside of the resonator, it is impossible to determine whether it was the first, the second and so on. Naturally, the question arises whether effects observed on single trajectories (like entanglement) ``survive'' this averaging and can still be expected to be observable after jumps. Every time a photon is detected, the mixed state inside the resonator is projected into the state
\begin{equation}\label{eq_rhoj}
  \rho_\mathrm{j}:=\frac{a\rho a^\dagger}{\Tr\left(a\rho a^\dagger\right)},
\end{equation}
where $\rho$ is the density matrix evolved according to the master equation~\eqref{eq_master}. Interestingly, the momentum correlation coefficient~\eqref{eq_Cp} is nearly constant ($\Cp\approx 0.5$ for $\DC=-\kappa$ and $\Cp\approx 0.3$ on the cooling sideband, respectively), regardless of the time the jump occurred at, see figure~\ref{fig_rhoj} also. The heralded entanglement measured by the logarithmic negativity is smaller than on single trajectories, but still prominent with a value $\EN\sim 0.2$ for our parameters and a jump in steady state.

\par
The conditional density matrix~\eqref{eq_rhoj} is closely related to the field autocorrelation function~\cite{gardinerbook}
\begin{equation}
  g^{(2)}(0)=\frac{\ew{a^\dagger a^\dagger a a}}{\ew{a^\dagger a}^2}\equiv\frac{\ew{a^\dagger a}_{\rho_\mathrm{j}}}{\ew{a^\dagger a}_{\rho}}.
\end{equation}
It has the very intuitive interpretation as the ratio between the photon number after and prior to a jump in steady state~\cite{alge1997nondegenerate}. For our parameters, it indicates photon bunching close to a thermal (chaotic) state (for which $g^{(2)}(0)=2$). This is consistent with the picture of the mode being incoherently populated through particle noise---perfectly localized particles do not scatter. Hence the field expectation vanishes and only its variance gives a contribution.

\section{Gaussian optomechanical treatment}

So far, we have studied the mainly numerically accessible general system~\eqref{eq_H} of two particles in a ring resonator. Now, we investigate the special case of tightly confined particles allowing for more analytical insight into the dynamics. In this limit, it is justified to keep only the first- and second-order terms in the expansion of the trigonometric factors and the Hamiltonian~\eqref{eq_H} can be mapped onto the linearized optomechanical model
\begin{equation}\label{eq_H_ho}
  H=\sum_{i=1}^2{\hbar\omega b_i^\dagger b_i}-\hbar\DC a^\dagger a + \hbar g \sum_{i=1}^2(b_i+b_i^\dagger)(a+a^\dagger)
\end{equation}
as shown in~\cite{schulze2010optomechanical}. Here we have defined $\omega:=\sqrt{4\hbar|U_0|\alphac^2\Erec}/\hbar$, $g:=U_0\alphac k\xi_0/\sqrt{2}$, the oscillator length $\xi_0:=\sqrt{\hbar/m\omega}$ and $b_i:=(x_i/\xi_0+\ii\xi_0p_i/\hbar)/\sqrt{2}$. The scaling of the oscillator length suggests that the Lamb--Dicke regime $k\xi_0\ll 1$ may also be reached for very heavy particles in shallower traps. As we do not consider any direct particle--particle interactions, the Hamiltonian~\eqref{eq_H_ho} is also valid for particles not confined in the same, but rather in distant sites within the resonator. 
\par
Interestingly, the Hamiltonian~\eqref{eq_H_ho} applies to a whole class of systems. A few well-studied realizations are (i) a cavity with vibrating end mirrors where the mirror--light interaction is always linear and governed mainly by the cavity length and the zero point motion of the mechanics~\cite{wilson2007theory,pinard2005entangling}, (ii) two light membranes inside a cavity field positioned at the maximum slope of the field amplitude where the coupling depends on the reflective properties of the membranes and is increased with decreasing mass~\cite{bhattacharya2008multiple} and (iii) very light nano-sized dielectric spheres held inside the cavity field either by an external trapping light mode or by means of optical tweezers~\cite{romero2011optically}.
\par
The Hamiltonian~\eqref{eq_H_ho} is quadratic in the bosonic operators so that an initially Gaussian state remains Gaussian throughout the time evolution. Gaussian states are completely defined by a displacement vector and a covariance matrix $V_{ij}:=\frac{1}{2}\left[\cov(\xi_i,\xi_j)+\cov(\xi_j,\xi_i)\right]$, with $\xi:=\left(\tilde{x}_1,\tilde{p}_1,\tilde{x}_2,\tilde{p}_2,X,P\right)$, $\tilde{x}_i:=x_i/\xi_0=(b_i+b^\dagger_i)/\sqrt{2}$, $\tilde{p}_i:=p_i\xi_0/\hbar=(b_i-b^\dagger_i)/(\ii\sqrt{2})$ and the field quadratures $X$ and $P$~\cite{adesso2007entanglement}. As shown in~\cite{carmichaelbook}, the time evolution of $V$ is determined by the equation (its steady-state version is called the Lyapunov equation)
\begin{equation}\label{eq_lyapunov}
  \dot V(t)=A V(t) + V(t)A^T + B,
\end{equation}
where $A$ and $B$ are the drift- and diffusion matrices appearing in stochastic Heisenberg--Langevin equations equivalent to the master equation. The steady-state solution of~\eqref{eq_lyapunov} for the covariances is $\cov(\tilde{p}_1,\tilde{p}_2)=\var(\tilde{p}_i)-1/2$ and $\cov(\tilde{x}_1,\tilde{x}_2)=\var(\tilde{x}_i)-1/2$ (both particles behave the same way). It only exists in the cooling regime $\DC<0$ since the momentum variance is only positive for red detuning. The momentum covariance is also genuinely positive. For $\Cp$ we find the simple ($g$-independent) expression
\begin{equation}
  \Cp=\frac{\kappa^2+(\DC+\omega)^2}{\kappa^2+(\DC-\omega)^2}>0,
\end{equation}
which is precisely the ratio of the Stokes- ($\Gamma_-$, heating) and anti-Stokes ($\Gamma_+$, cooling) scattering rates found when adiabatically eliminating the cavity field~\cite{wallquist2010single}. These rates $\Gamma_\pm=\frac{g^2\kappa}{\kappa^2+(\DC\pm\omega)^2}$ also define the time scale $\tau\sim(\Gamma_+-\Gamma_-)^{-1}$ on which the steady state is reached.
\par
\begin{figure}
  \centering
  \includegraphics[width=0.8\columnwidth]{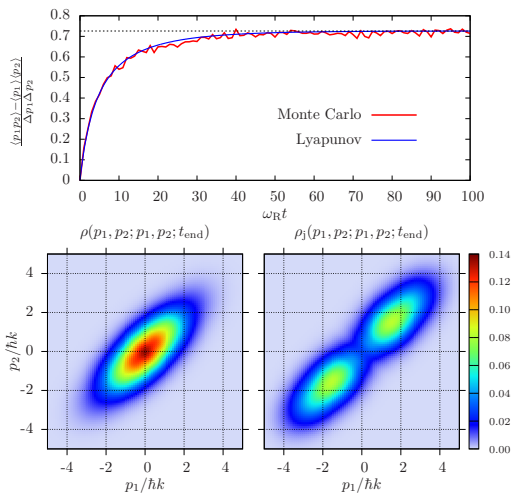}
  \caption{Upper plot: momentum correlation obtained from Monte Carlo simulations in the Lamb--Dicke limit for $k\xi_0=0.1$ (red, ensemble of 500 trajectories) and solution of equation~\eqref{eq_lyapunov} (blue). The position correlation is likewise pronounced. Lower plots: diagonal elements of the reduced particle density matrix in steady state (left) and of the reduced conditional density matrix~\eqref{eq_rhoj} (right). Parameters: $\omega=200\omrec$, $g=5\omrec$, $\kappa=100\omrec$ and $\DC=-20\omrec$. Note that here the particles--field interaction is much stronger than in figure~\ref{fig_singletrajectory} where $g\approx0.2\omrec$.}\label{fig_momentumcorrelation_lyapunov}
\end{figure}

$\Cp$ has a minimum at $\DC=-\sqrt{\kappa^2+\omega^2}$ and approaches unity for $|\DC|\rightarrow\infty$ and $\DC\rightarrow 0^-$. From the form of the steady-state covariance matrix we can deduce that steady-state entanglement between the particles can only occur if $\cov(\tilde{x}_1,\tilde{x}_2)<0$, which implies squeezing of the position variable. This is a direct consequence of the entanglement criterion derived in~\cite{simon2000peres}. See figure~\ref{fig_momentumcorrelation_lyapunov} for an example of the Lyapunov time evolution. For such deep potentials (a regime which we can numerically analyse by changing the effective parameters $\omega$ and $g$ to the values listed in figure~\ref{fig_momentumcorrelation_lyapunov}) we restricted ourselves to Monte Carlo simulations of the master equation containing the Hamiltonian~\eqref{eq_H_ho}. Again, the C++QED framework provides a helpful basis for their numerical implementation as it also supports the simulation of coupled oscillators.
\par
The steady-state covariance matrix reveals strong correlations between the particles and the field quadratures, making it intuitively clear that the reduced conditional density matrix for the particles strongly differs from its steady-state counterpart. Each photon detection highly influences the joint state of particles and cavity field. We show both density matrices in figure~\ref{fig_momentumcorrelation_lyapunov}. The logarithmic negativity for the conditional density matrix is $\EN\approx0.25$ and $\Cp$ is found to be $\Cp\approx 0.9$. Due to the particles--field correlations in steady state the conditional density matrix~\eqref{eq_rhoj} does not describe a Gaussian state. Indeed, it reveals a double-peak structure.

Given that steady state dynamics shows little or no traces of the entanglement that is clearly present in individual trajectories, we pursue now an alternative road that investigates the transient regime. Such a regime could for example be reached with short light pulses that can be used to generate build-up of motional correlations in the system. While we are free to explore any regime numerically, the problem can be analytically tackled mainly in the adiabatic case where the cavity mode can be eliminated from the dynamics, when $g\ll\kappa$ or $g\ll|\DC\pm\omega|$~\cite{wallquist2010single}. We verify, however, that the generality is not lost since numerical investigations show the optimal entanglement regime indeed being around the point analytically treated.
\par
\begin{figure}
  \centering
  \includegraphics[width=0.8\columnwidth]{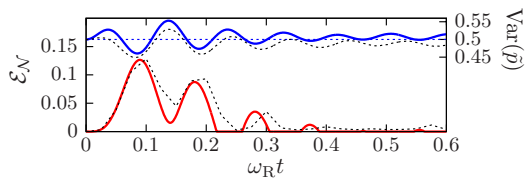}
  \caption{Transient entanglement (lower red curve) and momentum variance (upper blue curve) as obtained from the Lyapunov equation. Black dashed lines: solution of the master equation for the ring resonator (2000 trajectories), for comparison. Parameters: $\omega=30\omrec$, $g=5\omrec$, $\kappa=5\omrec$ and $\DC=-(\omega-\kappa)$.}\label{fig_transient_entanglement}
\end{figure}

We skip the cumbersome analytical elimination procedure (that closely follows the one used in~\cite{wallquist2010single}) and simply make use of the simple form of the end result that shows displacement--displacement coupling of the reduced bipartite system, $H_\mathrm{eff}^\mathrm{int} \approx -\hbar\Upsilon\sum_{jk}\tilde{x}_j\tilde{x}_k$, with 
\begin{equation}
  \Upsilon=-\left(\frac{\DC-\omega}{\kappa}\Gamma_-+\frac{\DC+\omega}{\kappa}\Gamma_+\right).
\end{equation}
A second result of such an elimination procedure is that decay of the system can occur in a correlated fashion, namely with an effective jump operator proportional to $b_1+b_2$. Now, we choose the operating conditions such that $\tilde{x}_1\tilde{x}_2$ reduces to a simple beam splitter interaction $b_1^\dagger b_2+b_2^\dagger b_1$ and point out a very general conclusion (see~\cite{kim2002entanglement}) that this interaction does not lead to entanglement when starting in non-correlated initial states but rather to a simple state swap process where in the limit of strong coupling (coupling larger than decoherence rates), a quantum state of a subsystem can be written on the other subsystem. Than, taking advantage of the correlated decay, we show that transient entanglement is present in transient dynamics. This is illustrated in figure~\ref{fig_transient_entanglement} where we show the time evolution of the logarithmic negativity. The relatively large entanglement obtained at half of the interaction time (defined by the inverse of the effective coupling strength $g$) can be accessed in an experimental situation for example by the use of laser pulses of tailored shape and duration.

Our effective optomechanical Hamiltonian~\eqref{eq_H_ho} is formally similar to a model describing the microscopic dynamics of self-organization of particles in a cavity field~\cite{vukics2007microscopic}. The behaviour of single trajectories there can be explained with the help of an adiabatic model. We now strive to develop a model in the spirit of~\cite{vukics2007microscopic}, capable of describing the behaviour of single Monte Carlo trajectories of the master equation containing the Hamiltonian~\eqref{eq_H_ho}. To this aim, we diagonalize the scattering operator $\sum_i(b_i+b_i^\dagger)$ to obtain the new particle basis $\{\ket{e_i}\}$ containing radiative (eigenvalue $\lambda_i\neq 0$, they always appear with both signs $\pm1$) and non-radiative ($\lambda_i=0$) states. Ignoring terms stemming from the oscillator energies for the moment, the radiative states radiate a field $\ket{\lambda_i\alpha}$ with $\alpha:=-\ii g/(\kappa-\ii\DC)$. The crucial assumption now is adiabaticity (cavity decay assumed to define the fastest time scale), \ie we assume each particle state $\ket{e_i}$ to be correlated with its associated unperturbed field state to find the approximated joint particle--field stochastic state vector
\begin{equation}\label{eq_psi_stoch}
  \ket{\psi(t)}=\sum_jc_j(t)\ket{e_j}\ket{\lambda_j\alpha}.
\end{equation}
The coefficients $c_i(t)$ are determined by the effective conditional Monte Carlo time evolution generated by the non-Hermitian Hamiltonian (abbreviating $\ket{i}:=\ket{e_i}$)
\begin{equation}\label{eq_H_toy}
  H_\mathrm{nH}:=\hbar\omega\sum_{ij}\bra{i}\sum_kb_k^\dagger b_k\ket{j}\ketbra{i}{j}-\ii\hbar\kappa\sum_i|\lambda_i\alpha|^2\proj{i}
\end{equation}
for the particles only. Its Hermitian part contains the oscillator energies expressed in the scattering basis and couples the constituents of the latter. 
\par
From the form of the state vector~\eqref{eq_psi_stoch} one can conclude what happens if a photon is detected. Applying the jump operator $\propto a$ on the internal state~\eqref{eq_psi_stoch} at time $t_\mathrm{j}$ yields $a\ket{\psi(t_\mathrm{j})}\propto \sum_i c_i(t_\mathrm{j})\lambda_i\ket{e_i}\ket{\lambda_i\alpha}$. Hence, all information about the non-radiative contributions are erased and the relative phases between the radiative states change sign (as for each $\lambda_i$ there exists also an eigenvalue $-\lambda_i$). Since $|\lambda_i\alpha|^2\ll1$ the reduced density matrix for the particles directly after a jump mainly contains the pure state $\ket{\psi_\mathrm{p}}\propto\sum_j c_j(t)\lambda_j\ket{e_j}$ (during the time evolution the particle state reads $\ket{\psi_\mathrm{p}}\approx\sum_j c_j(t)\ket{e_j}$). Inserting the coefficients for a state describing two particles in the ground state, the first jump excites one of them and the photon number increases by a factor of 3. The resulting state of the particles $\ket{\psi_\mathrm{p}}\approx\frac{1}{\sqrt{2}}\left(\ket{\psi_0\psi_1}+\ket{\psi_1\psi_0}\right)$ is entangled with $\EN=1$ and correlated with $\Cp=1/2$. Beginning with the second jump, fast oscillations with a frequency determined by $\omega$ build up, which is in qualitative agreement with single trajectories of the ring resonator model as shown in figure~\ref{fig_singletrajectory}. This way it is possible to qualitatively and quantitatively reconstruct single trajectories of the optomechanical model~\eqref{eq_H_ho}.

\section{Conclusion and outlook}

Starting from two particles in a ring resonator, our conclusions are quite general. We have proposed a tripartite optomechanical system with two identical oscillators that can be correlated via the mediation of a photon field and the interaction amplified by the cavity confinement. The Gaussian evolution of the reduced mechanical bipartition has been followed both numerically and analytically, while non-Gaussian effects owing to the deviation from the bilinear Hamiltonian have been investigated numerically. The non-intuitive and classically forbidden positive momentum correlations are an important result of the paper, and are interpreted as strong delocalizations (superpositions) of the centre-of-mass independent of the particle separation and thus of the extension of the effective system; this suggests possible use of the setup to test the quantum-classical boundary. Furthermore, we have studied entanglement on single Monte Carlo trajectories as well as of the averaged density matrix. Entanglement heralded by photodetection has also been investigated. Transient entanglement as shown here can also be exploited in a pulsed regime where light pulses can drive the mechanical bipartition into a strongly entangled state.

\begin{acknowledgments}
This work has been supported by the Austrian Science Fund FWF through projects P20391, P207481 and F4013 and by the EU-ITN CCQED project. We would like to thank Cecilia Cormick, Tobias Grie{\ss}er, Sebastian~Kr\"amer and Andr\'as Vukics for helpful discussions and Hans Embacher for technical support.
\end{acknowledgments}

%\bibliographystyle{apsrev}
%\bibliography{ringcorr}

\begin{thebibliography}{38}
\expandafter\ifx\csname natexlab\endcsname\relax\def\natexlab#1{#1}\fi
\expandafter\ifx\csname bibnamefont\endcsname\relax
  \def\bibnamefont#1{#1}\fi
\expandafter\ifx\csname bibfnamefont\endcsname\relax
  \def\bibfnamefont#1{#1}\fi
\expandafter\ifx\csname citenamefont\endcsname\relax
  \def\citenamefont#1{#1}\fi
\expandafter\ifx\csname url\endcsname\relax
  \def\url#1{\texttt{#1}}\fi
\expandafter\ifx\csname urlprefix\endcsname\relax\def\urlprefix{URL }\fi
\providecommand{\bibinfo}[2]{#2}
\providecommand{\eprint}[2][]{\url{#2}}

\bibitem[{\citenamefont{Domokos and Ritsch}(2003)}]{domokos2003mechanical}
\bibinfo{author}{\bibfnamefont{P.}~\bibnamefont{Domokos}} \bibnamefont{and}
  \bibinfo{author}{\bibfnamefont{H.}~\bibnamefont{Ritsch}},
  \bibinfo{journal}{J. Opt. Soc. Am. B} \textbf{\bibinfo{volume}{20}},
  \bibinfo{pages}{1098} (\bibinfo{year}{2003}).

\bibitem[{\citenamefont{Mekhov and Ritsch}(2012)}]{mekhov2012quantum}
\bibinfo{author}{\bibfnamefont{I.~B.} \bibnamefont{Mekhov}} \bibnamefont{and}
  \bibinfo{author}{\bibfnamefont{H.}~\bibnamefont{Ritsch}},
  \bibinfo{journal}{J. Phys. B} \textbf{\bibinfo{volume}{45}},
  \bibinfo{pages}{102001} (\bibinfo{year}{2012}).

\bibitem[{\citenamefont{Gupta et~al.}(2007)\citenamefont{Gupta, Moore, Murch,
  and Stamper-Kurn}}]{gupta2007cavity}
\bibinfo{author}{\bibfnamefont{S.}~\bibnamefont{Gupta}},
  \bibinfo{author}{\bibfnamefont{K.~L.} \bibnamefont{Moore}},
  \bibinfo{author}{\bibfnamefont{K.~W.} \bibnamefont{Murch}}, \bibnamefont{and}
  \bibinfo{author}{\bibfnamefont{D.~M.} \bibnamefont{Stamper-Kurn}},
  \bibinfo{journal}{Phys. Rev. Lett.} \textbf{\bibinfo{volume}{99}},
  \bibinfo{pages}{213601} (\bibinfo{year}{2007}).

\bibitem[{\citenamefont{Leroux et~al.}(2010)\citenamefont{Leroux,
  Schleier-Smith, and Vuleti\'{c}}}]{leroux2010implementation}
\bibinfo{author}{\bibfnamefont{I.~D.} \bibnamefont{Leroux}},
  \bibinfo{author}{\bibfnamefont{M.~H.} \bibnamefont{Schleier-Smith}},
  \bibnamefont{and}
  \bibinfo{author}{\bibfnamefont{V.}~\bibnamefont{Vuleti\'{c}}},
  \bibinfo{journal}{Phys. Rev. Lett.} \textbf{\bibinfo{volume}{104}},
  \bibinfo{pages}{073602} (\bibinfo{year}{2010}).

\bibitem[{\citenamefont{Eschner et~al.}(2003)\citenamefont{Eschner, Morigi,
  Schmidt-Kaler, and Blatt}}]{eschner2003laser}
\bibinfo{author}{\bibfnamefont{J.}~\bibnamefont{Eschner}},
  \bibinfo{author}{\bibfnamefont{G.}~\bibnamefont{Morigi}},
  \bibinfo{author}{\bibfnamefont{F.}~\bibnamefont{Schmidt-Kaler}},
  \bibnamefont{and} \bibinfo{author}{\bibfnamefont{R.}~\bibnamefont{Blatt}},
  \bibinfo{journal}{J. Opt. Soc. Am. B} \textbf{\bibinfo{volume}{20}},
  \bibinfo{pages}{1003} (\bibinfo{year}{2003}).

\bibitem[{\citenamefont{Thompson et~al.}(2008)\citenamefont{Thompson, Zwickl,
  Jayich, Marquardt, Girvin, and Harris}}]{thompson2008strong}
\bibinfo{author}{\bibfnamefont{J.~D.} \bibnamefont{Thompson}},
  \bibinfo{author}{\bibfnamefont{B.~M.} \bibnamefont{Zwickl}},
  \bibinfo{author}{\bibfnamefont{A.~M.} \bibnamefont{Jayich}},
  \bibinfo{author}{\bibfnamefont{F.}~\bibnamefont{Marquardt}},
  \bibinfo{author}{\bibfnamefont{S.~M.} \bibnamefont{Girvin}},
  \bibnamefont{and} \bibinfo{author}{\bibfnamefont{J.~G.~E.}
  \bibnamefont{Harris}}, \bibinfo{journal}{Nature}
  \textbf{\bibinfo{volume}{452}}, \bibinfo{pages}{72} (\bibinfo{year}{2008}).

\bibitem[{\citenamefont{Gr{\"o}blacher
  et~al.}(2009)\citenamefont{Gr{\"o}blacher, Hertzberg, Vanner, Cole, Gigan,
  Schwab, and Aspelmeyer}}]{groeblacher2009demonstration}
\bibinfo{author}{\bibfnamefont{S.}~\bibnamefont{Gr{\"o}blacher}},
  \bibinfo{author}{\bibfnamefont{J.~B.} \bibnamefont{Hertzberg}},
  \bibinfo{author}{\bibfnamefont{M.~R.} \bibnamefont{Vanner}},
  \bibinfo{author}{\bibfnamefont{G.~D.} \bibnamefont{Cole}},
  \bibinfo{author}{\bibfnamefont{S.}~\bibnamefont{Gigan}},
  \bibinfo{author}{\bibfnamefont{K.~C.} \bibnamefont{Schwab}},
  \bibnamefont{and}
  \bibinfo{author}{\bibfnamefont{M.}~\bibnamefont{Aspelmeyer}},
  \bibinfo{journal}{Nature Phys.} \textbf{\bibinfo{volume}{5}},
  \bibinfo{pages}{485} (\bibinfo{year}{2009}).

\bibitem[{\citenamefont{Romero-Isart et~al.}(2011)\citenamefont{Romero-Isart,
  Pflanzer, Juan, Quidant, Kiesel, Aspelmeyer, and
  Cirac}}]{romero2011optically}
\bibinfo{author}{\bibfnamefont{O.}~\bibnamefont{Romero-Isart}},
  \bibinfo{author}{\bibfnamefont{A.~C.} \bibnamefont{Pflanzer}},
  \bibinfo{author}{\bibfnamefont{M.~L.} \bibnamefont{Juan}},
  \bibinfo{author}{\bibfnamefont{R.}~\bibnamefont{Quidant}},
  \bibinfo{author}{\bibfnamefont{N.}~\bibnamefont{Kiesel}},
  \bibinfo{author}{\bibfnamefont{M.}~\bibnamefont{Aspelmeyer}},
  \bibnamefont{and} \bibinfo{author}{\bibfnamefont{J.~I.} \bibnamefont{Cirac}},
  \bibinfo{journal}{Phys. Rev. A} \textbf{\bibinfo{volume}{83}},
  \bibinfo{pages}{013803} (\bibinfo{year}{2011}).

\bibitem[{\citenamefont{Wallquist et~al.}(2010)\citenamefont{Wallquist,
  Hammerer, Zoller, Genes, Ludwig, Marquardt, Treutlein, Ye, and
  Kimble}}]{wallquist2010single}
\bibinfo{author}{\bibfnamefont{M.}~\bibnamefont{Wallquist}},
  \bibinfo{author}{\bibfnamefont{K.}~\bibnamefont{Hammerer}},
  \bibinfo{author}{\bibfnamefont{P.}~\bibnamefont{Zoller}},
  \bibinfo{author}{\bibfnamefont{C.}~\bibnamefont{Genes}},
  \bibinfo{author}{\bibfnamefont{M.}~\bibnamefont{Ludwig}},
  \bibinfo{author}{\bibfnamefont{F.}~\bibnamefont{Marquardt}},
  \bibinfo{author}{\bibfnamefont{P.}~\bibnamefont{Treutlein}},
  \bibinfo{author}{\bibfnamefont{J.}~\bibnamefont{Ye}}, \bibnamefont{and}
  \bibinfo{author}{\bibfnamefont{H.~J.} \bibnamefont{Kimble}},
  \bibinfo{journal}{Phys. Rev. A} \textbf{\bibinfo{volume}{81}},
  \bibinfo{pages}{023816} (\bibinfo{year}{2010}).

\bibitem[{\citenamefont{Nunnenkamp et~al.}(2010)\citenamefont{Nunnenkamp,
  B\o{}rkje, Harris, and Girvin}}]{nunnenkamp2010cooling}
\bibinfo{author}{\bibfnamefont{A.}~\bibnamefont{Nunnenkamp}},
  \bibinfo{author}{\bibfnamefont{K.}~\bibnamefont{B\o{}rkje}},
  \bibinfo{author}{\bibfnamefont{J.~G.~E.} \bibnamefont{Harris}},
  \bibnamefont{and} \bibinfo{author}{\bibfnamefont{S.~M.}
  \bibnamefont{Girvin}}, \bibinfo{journal}{Phys. Rev. A}
  \textbf{\bibinfo{volume}{82}}, \bibinfo{pages}{021806}
  (\bibinfo{year}{2010}).

\bibitem[{\citenamefont{Krauter et~al.}(2011)\citenamefont{Krauter, Muschik,
  Jensen, Wasilewski, Petersen, Cirac, and Polzik}}]{krauter2011entanglement}
\bibinfo{author}{\bibfnamefont{H.}~\bibnamefont{Krauter}},
  \bibinfo{author}{\bibfnamefont{C.~A.} \bibnamefont{Muschik}},
  \bibinfo{author}{\bibfnamefont{K.}~\bibnamefont{Jensen}},
  \bibinfo{author}{\bibfnamefont{W.}~\bibnamefont{Wasilewski}},
  \bibinfo{author}{\bibfnamefont{J.~M.} \bibnamefont{Petersen}},
  \bibinfo{author}{\bibfnamefont{J.~I.} \bibnamefont{Cirac}}, \bibnamefont{and}
  \bibinfo{author}{\bibfnamefont{E.~S.} \bibnamefont{Polzik}},
  \bibinfo{journal}{Phys. Rev. Lett.} \textbf{\bibinfo{volume}{107}},
  \bibinfo{pages}{080503} (\bibinfo{year}{2011}).

\bibitem[{\citenamefont{Nha and Carmichael}(2004)}]{nha2004entanglement}
\bibinfo{author}{\bibfnamefont{H.}~\bibnamefont{Nha}} \bibnamefont{and}
  \bibinfo{author}{\bibfnamefont{H.~J.} \bibnamefont{Carmichael}},
  \bibinfo{journal}{Phys. Rev. Lett.} \textbf{\bibinfo{volume}{93}},
  \bibinfo{pages}{120408} (\bibinfo{year}{2004}).

\bibitem[{\citenamefont{Plenio et~al.}(1999)\citenamefont{Plenio, Huelga,
  Beige, and Knight}}]{plenio1999cavity}
\bibinfo{author}{\bibfnamefont{M.~B.} \bibnamefont{Plenio}},
  \bibinfo{author}{\bibfnamefont{S.~F.} \bibnamefont{Huelga}},
  \bibinfo{author}{\bibfnamefont{A.}~\bibnamefont{Beige}}, \bibnamefont{and}
  \bibinfo{author}{\bibfnamefont{P.~L.} \bibnamefont{Knight}},
  \bibinfo{journal}{Phys. Rev. A} \textbf{\bibinfo{volume}{59}},
  \bibinfo{pages}{2468} (\bibinfo{year}{1999}).

\bibitem[{\citenamefont{Abdi et~al.}(2011)\citenamefont{Abdi, Barzanjeh,
  Tombesi, and Vitali}}]{abdi2011effect}
\bibinfo{author}{\bibfnamefont{M.}~\bibnamefont{Abdi}},
  \bibinfo{author}{\bibfnamefont{S.}~\bibnamefont{Barzanjeh}},
  \bibinfo{author}{\bibfnamefont{P.}~\bibnamefont{Tombesi}}, \bibnamefont{and}
  \bibinfo{author}{\bibfnamefont{D.}~\bibnamefont{Vitali}},
  \bibinfo{journal}{Phys. Rev. A} \textbf{\bibinfo{volume}{84}},
  \bibinfo{pages}{032325} (\bibinfo{year}{2011}).

\bibitem[{\citenamefont{B{\o}rkje et~al.}(2011)\citenamefont{B{\o}rkje,
  Nunnenkamp, and Girvin}}]{boerkje2011proposal}
\bibinfo{author}{\bibfnamefont{K.}~\bibnamefont{B{\o}rkje}},
  \bibinfo{author}{\bibfnamefont{A.}~\bibnamefont{Nunnenkamp}},
  \bibnamefont{and} \bibinfo{author}{\bibfnamefont{S.~M.}
  \bibnamefont{Girvin}}, \bibinfo{journal}{Phys. Rev. Lett.}
  \textbf{\bibinfo{volume}{107}}, \bibinfo{pages}{123601}
  (\bibinfo{year}{2011}).

\bibitem[{\citenamefont{Joshi et~al.}(2012)\citenamefont{Joshi, Larson, Jonson,
  Andersson, and \"Ohberg}}]{joshi2012entanglement}
\bibinfo{author}{\bibfnamefont{C.}~\bibnamefont{Joshi}},
  \bibinfo{author}{\bibfnamefont{J.}~\bibnamefont{Larson}},
  \bibinfo{author}{\bibfnamefont{M.}~\bibnamefont{Jonson}},
  \bibinfo{author}{\bibfnamefont{E.}~\bibnamefont{Andersson}},
  \bibnamefont{and} \bibinfo{author}{\bibfnamefont{P.}~\bibnamefont{\"Ohberg}},
  \bibinfo{journal}{Phys. Rev. A} \textbf{\bibinfo{volume}{85}},
  \bibinfo{pages}{033805} (\bibinfo{year}{2012}).

\bibitem[{\citenamefont{Mancini and Tombesi}(2003)}]{mancini2003high}
\bibinfo{author}{\bibfnamefont{S.}~\bibnamefont{Mancini}} \bibnamefont{and}
  \bibinfo{author}{\bibfnamefont{P.}~\bibnamefont{Tombesi}},
  \bibinfo{journal}{Europhys. Lett.} \textbf{\bibinfo{volume}{61}},
  \bibinfo{pages}{8} (\bibinfo{year}{2003}).

\bibitem[{\citenamefont{Mancini et~al.}(2002)\citenamefont{Mancini,
  Giovannetti, Vitali, and Tombesi}}]{mancini2002entangling}
\bibinfo{author}{\bibfnamefont{S.}~\bibnamefont{Mancini}},
  \bibinfo{author}{\bibfnamefont{V.}~\bibnamefont{Giovannetti}},
  \bibinfo{author}{\bibfnamefont{D.}~\bibnamefont{Vitali}}, \bibnamefont{and}
  \bibinfo{author}{\bibfnamefont{P.}~\bibnamefont{Tombesi}},
  \bibinfo{journal}{Phys. Rev. Lett.} \textbf{\bibinfo{volume}{88}},
  \bibinfo{pages}{120401} (\bibinfo{year}{2002}).

\bibitem[{\citenamefont{Hartmann and Plenio}(2008)}]{hartmann2008steady}
\bibinfo{author}{\bibfnamefont{M.~J.} \bibnamefont{Hartmann}} \bibnamefont{and}
  \bibinfo{author}{\bibfnamefont{M.~B.} \bibnamefont{Plenio}},
  \bibinfo{journal}{Phys. Rev. Lett.} \textbf{\bibinfo{volume}{101}},
  \bibinfo{pages}{200503} (\bibinfo{year}{2008}).

\bibitem[{\citenamefont{Schulze et~al.}(2010)\citenamefont{Schulze, Genes, and
  Ritsch}}]{schulze2010optomechanical}
\bibinfo{author}{\bibfnamefont{R.~J.} \bibnamefont{Schulze}},
  \bibinfo{author}{\bibfnamefont{C.}~\bibnamefont{Genes}}, \bibnamefont{and}
  \bibinfo{author}{\bibfnamefont{H.}~\bibnamefont{Ritsch}},
  \bibinfo{journal}{Phys. Rev. A} \textbf{\bibinfo{volume}{81}},
  \bibinfo{pages}{063820} (\bibinfo{year}{2010}).

\bibitem[{\citenamefont{Niedenzu et~al.}(2010)\citenamefont{Niedenzu, Schulze,
  Vukics, and Ritsch}}]{niedenzu2010microscopic}
\bibinfo{author}{\bibfnamefont{W.}~\bibnamefont{Niedenzu}},
  \bibinfo{author}{\bibfnamefont{R.}~\bibnamefont{Schulze}},
  \bibinfo{author}{\bibfnamefont{A.}~\bibnamefont{Vukics}}, \bibnamefont{and}
  \bibinfo{author}{\bibfnamefont{H.}~\bibnamefont{Ritsch}},
  \bibinfo{journal}{Phys. Rev. A} \textbf{\bibinfo{volume}{82}},
  \bibinfo{pages}{043605} (\bibinfo{year}{2010}).

\bibitem[{\citenamefont{Gangl and Ritsch}(2000)}]{gangl2000cold}
\bibinfo{author}{\bibfnamefont{M.}~\bibnamefont{Gangl}} \bibnamefont{and}
  \bibinfo{author}{\bibfnamefont{H.}~\bibnamefont{Ritsch}},
  \bibinfo{journal}{Phys. Rev. A} \textbf{\bibinfo{volume}{61}},
  \bibinfo{pages}{043405} (\bibinfo{year}{2000}).

\bibitem[{\citenamefont{Gardiner and Zoller}(2000)}]{gardinerbook}
\bibinfo{author}{\bibfnamefont{C.~W.} \bibnamefont{Gardiner}} \bibnamefont{and}
  \bibinfo{author}{\bibfnamefont{P.}~\bibnamefont{Zoller}},
  \emph{\bibinfo{title}{Quantum noise}} (\bibinfo{publisher}{Springer-Verlag},
  \bibinfo{address}{Berlin}, \bibinfo{year}{2000}), \bibinfo{edition}{2nd} ed.

\bibitem[{\citenamefont{M{\o}lmer et~al.}(1993)\citenamefont{M{\o}lmer, Castin,
  and Dalibard}}]{moelmer1993monte}
\bibinfo{author}{\bibfnamefont{K.}~\bibnamefont{M{\o}lmer}},
  \bibinfo{author}{\bibfnamefont{Y.}~\bibnamefont{Castin}}, \bibnamefont{and}
  \bibinfo{author}{\bibfnamefont{J.}~\bibnamefont{Dalibard}},
  \bibinfo{journal}{J. Opt. Soc. Am. B} \textbf{\bibinfo{volume}{10}},
  \bibinfo{pages}{524} (\bibinfo{year}{1993}).

\bibitem[{\citenamefont{Vukics and Ritsch}(2007)}]{vukics2007object}
\bibinfo{author}{\bibfnamefont{A.}~\bibnamefont{Vukics}} \bibnamefont{and}
  \bibinfo{author}{\bibfnamefont{H.}~\bibnamefont{Ritsch}},
  \bibinfo{journal}{Eur. Phys. J. D} \textbf{\bibinfo{volume}{44}},
  \bibinfo{pages}{585} (\bibinfo{year}{2007}).

\bibitem[{\citenamefont{Vukics}(2012)}]{vukics2012cppqedv2}
\bibinfo{author}{\bibfnamefont{A.}~\bibnamefont{Vukics}},
  \bibinfo{journal}{Computer Physics Communications}
  \textbf{\bibinfo{volume}{183}}, \bibinfo{pages}{1381} (\bibinfo{year}{2012}).

\bibitem[{\citenamefont{Vidal and Werner}(2002)}]{vidal2002computable}
\bibinfo{author}{\bibfnamefont{G.}~\bibnamefont{Vidal}} \bibnamefont{and}
  \bibinfo{author}{\bibfnamefont{R.~F.} \bibnamefont{Werner}},
  \bibinfo{journal}{Phys. Rev. A} \textbf{\bibinfo{volume}{65}},
  \bibinfo{pages}{032314} (\bibinfo{year}{2002}).

\bibitem[{\citenamefont{Gangl et~al.}(2000)\citenamefont{Gangl, Horak, and
  Ritsch}}]{gangl2000cooling}
\bibinfo{author}{\bibfnamefont{M.}~\bibnamefont{Gangl}},
  \bibinfo{author}{\bibfnamefont{P.}~\bibnamefont{Horak}}, \bibnamefont{and}
  \bibinfo{author}{\bibfnamefont{H.}~\bibnamefont{Ritsch}},
  \bibinfo{journal}{J. Mod. Opt.} \textbf{\bibinfo{volume}{47}},
  \bibinfo{pages}{2741} (\bibinfo{year}{2000}).

\bibitem[{\citenamefont{Sandner et~al.}(2012)}]{sandner2012}
\bibinfo{author}{\bibfnamefont{R.~M.} \bibnamefont{Sandner}}
  \bibnamefont{et~al.}, \bibinfo{journal}{in preparation}
  (\bibinfo{year}{2012}).

\bibitem[{\citenamefont{Alge et~al.}(1997)\citenamefont{Alge, Gheri, and
  Marte}}]{alge1997nondegenerate}
\bibinfo{author}{\bibfnamefont{W.}~\bibnamefont{Alge}},
  \bibinfo{author}{\bibfnamefont{K.~M.} \bibnamefont{Gheri}}, \bibnamefont{and}
  \bibinfo{author}{\bibfnamefont{M.~A.~M.} \bibnamefont{Marte}},
  \bibinfo{journal}{J. Mod. Opt.} \textbf{\bibinfo{volume}{44}},
  \bibinfo{pages}{841} (\bibinfo{year}{1997}).

\bibitem[{\citenamefont{Wilson-Rae et~al.}(2007)\citenamefont{Wilson-Rae,
  Nooshi, Zwerger, and Kippenberg}}]{wilson2007theory}
\bibinfo{author}{\bibfnamefont{I.}~\bibnamefont{Wilson-Rae}},
  \bibinfo{author}{\bibfnamefont{N.}~\bibnamefont{Nooshi}},
  \bibinfo{author}{\bibfnamefont{W.}~\bibnamefont{Zwerger}}, \bibnamefont{and}
  \bibinfo{author}{\bibfnamefont{T.~J.} \bibnamefont{Kippenberg}},
  \bibinfo{journal}{Phys. Rev. Lett.} \textbf{\bibinfo{volume}{99}},
  \bibinfo{pages}{093901} (\bibinfo{year}{2007}).

\bibitem[{\citenamefont{Pinard et~al.}(2005)\citenamefont{Pinard, Dantan,
  Vitali, Arcizet, Briant, and Heidmann}}]{pinard2005entangling}
\bibinfo{author}{\bibfnamefont{M.}~\bibnamefont{Pinard}},
  \bibinfo{author}{\bibfnamefont{A.}~\bibnamefont{Dantan}},
  \bibinfo{author}{\bibfnamefont{D.}~\bibnamefont{Vitali}},
  \bibinfo{author}{\bibfnamefont{O.}~\bibnamefont{Arcizet}},
  \bibinfo{author}{\bibfnamefont{T.}~\bibnamefont{Briant}}, \bibnamefont{and}
  \bibinfo{author}{\bibfnamefont{A.}~\bibnamefont{Heidmann}},
  \bibinfo{journal}{EPL} \textbf{\bibinfo{volume}{72}}, \bibinfo{pages}{747}
  (\bibinfo{year}{2005}).

\bibitem[{\citenamefont{Bhattacharya and
  Meystre}(2008)}]{bhattacharya2008multiple}
\bibinfo{author}{\bibfnamefont{M.}~\bibnamefont{Bhattacharya}}
  \bibnamefont{and} \bibinfo{author}{\bibfnamefont{P.}~\bibnamefont{Meystre}},
  \bibinfo{journal}{Phys. Rev. A} \textbf{\bibinfo{volume}{78}},
  \bibinfo{pages}{041801} (\bibinfo{year}{2008}).

\bibitem[{\citenamefont{Adesso and Illuminati}(2007)}]{adesso2007entanglement}
\bibinfo{author}{\bibfnamefont{G.}~\bibnamefont{Adesso}} \bibnamefont{and}
  \bibinfo{author}{\bibfnamefont{F.}~\bibnamefont{Illuminati}},
  \bibinfo{journal}{J. Phys. A: Math. Theor.} \textbf{\bibinfo{volume}{40}},
  \bibinfo{pages}{7821} (\bibinfo{year}{2007}).

\bibitem[{\citenamefont{Carmichael}(1999)}]{carmichaelbook}
\bibinfo{author}{\bibfnamefont{H.~J.} \bibnamefont{Carmichael}},
  \emph{\bibinfo{title}{Statistical Methods in Quantum Optics 1}}
  (\bibinfo{publisher}{Springer-Verlag}, \bibinfo{address}{Berlin},
  \bibinfo{year}{1999}).

\bibitem[{\citenamefont{Simon}(2000)}]{simon2000peres}
\bibinfo{author}{\bibfnamefont{R.}~\bibnamefont{Simon}},
  \bibinfo{journal}{Phys. Rev. Lett.} \textbf{\bibinfo{volume}{84}},
  \bibinfo{pages}{2726} (\bibinfo{year}{2000}).

\bibitem[{\citenamefont{Kim et~al.}(2002)\citenamefont{Kim, Son,
  Bu\ifmmode~\check{z}\else \v{z}\fi{}ek, and Knight}}]{kim2002entanglement}
\bibinfo{author}{\bibfnamefont{M.~S.} \bibnamefont{Kim}},
  \bibinfo{author}{\bibfnamefont{W.}~\bibnamefont{Son}},
  \bibinfo{author}{\bibfnamefont{V.}~\bibnamefont{Bu\ifmmode~\check{z}\else
  \v{z}\fi{}ek}}, \bibnamefont{and} \bibinfo{author}{\bibfnamefont{P.~L.}
  \bibnamefont{Knight}}, \bibinfo{journal}{Phys. Rev. A}
  \textbf{\bibinfo{volume}{65}}, \bibinfo{pages}{032323}
  (\bibinfo{year}{2002}).

\bibitem[{\citenamefont{Vukics et~al.}(2007)\citenamefont{Vukics, Maschler, and
  Ritsch}}]{vukics2007microscopic}
\bibinfo{author}{\bibfnamefont{A.}~\bibnamefont{Vukics}},
  \bibinfo{author}{\bibfnamefont{C.}~\bibnamefont{Maschler}}, \bibnamefont{and}
  \bibinfo{author}{\bibfnamefont{H.}~\bibnamefont{Ritsch}},
  \bibinfo{journal}{New J. Phys.} \textbf{\bibinfo{volume}{9}},
  \bibinfo{pages}{255} (\bibinfo{year}{2007}).

\end{thebibliography}

\end{document}